\documentclass[11pt,a4paper]{article}

\usepackage[scaled]{helvet} 
\usepackage[T1]{fontenc}

\usepackage{csquotes}
\usepackage[hidelinks]{hyperref}

\usepackage[style=ieee,backend=biber,maxnames=6,minnames=6]{biblatex}
\AtEveryBibitem{
  \clearfield{month} 
  \clearfield{day}   
  \clearfield{editor}
  \clearlist{editor}
}

\AtEveryBibitem{%
  \ifentrytype{article}{%
    \clearfield{abstract}%
    \clearfield{keywords}%
    \clearfield{issn}%
    \clearfield{isbn}%
    \clearfield{note}%
    \clearfield{file}%
    \clearfield{language}%
    \clearfield{series}%
    \clearfield{eprintclass}%
  }{}%

  \ifentrytype{inproceedings}{%
    \clearfield{isbn}%
    \clearfield{series}%
    \clearfield{address}%
    \clearfield{location}%
    \clearlist{location}%
    \clearfield{organization}%
    \clearfield{abstract}%
    \clearfield{keywords}%
    \clearfield{note}%
    \clearfield{file}%
    \clearfield{language}%
    \iffieldundef{url}{}{ %
      \clearfield{volume}%
      \clearlist{publisher}%
      \clearfield{publisher}%
    }{}%
  }{}%

  \ifentrytype{online}{%
    \clearfield{isbn}%
    \clearfield{abstract}%
    \clearfield{keywords}%
    \clearfield{publisher}%
    \clearfield{doi}%
    \clearfield{file}%
    \clearfield{language}%
  }{}%

  \ifentrytype{book}{%
    \clearfield{abstract}%
    \clearfield{keywords}%
    \clearfield{series}%
    \clearfield{isbn}%
    \clearfield{doi}%
    \clearfield{note}%
    \clearfield{file}%
    \clearfield{language}%
  }{}%

  \ifentrytype{incollection}{%
    \clearfield{isbn}%
    \clearfield{editor}%
    \clearfield{series}%
    \clearfield{address}%
    \clearfield{location}%
    \clearfield{organization}%
    \clearfield{abstract}%
    \clearfield{keywords}%
    \clearfield{note}%
    \clearfield{file}%
    \clearfield{language}%
  }{}%
}

\DeclareFieldFormat{doi}{%
  \ifhyperref
    {doi\addcolon\space\href{https://doi.org/#1}{\nolinkurl{#1}}}
    {doi\addcolon\space\nolinkurl{#1}}}

\usepackage{xurl}

\usepackage{graphicx}
\usepackage{xcolor}
\usepackage{lineno}
\usepackage{comment}
\usepackage{titlesec}
\usepackage{authblk}
\usepackage{tikz}
\usepackage{ragged2e}
\usepackage{fontawesome5}

\titlespacing*{\section}{0pt}{5pt}{-1pt} 
\titlespacing*{\subsection}{0pt}{5pt}{-1pt} 
\titlespacing*{\subsubsection}{0pt}{5pt}{-1pt}

\makeatletter
\newcommand{\addlogo}{
    \begin{tikzpicture}[remember picture,overlay]
        \node[anchor=north east, xshift=-1.75cm, yshift=-1cm] 
        at (current page.north east) {
            \includegraphics[width=6cm]{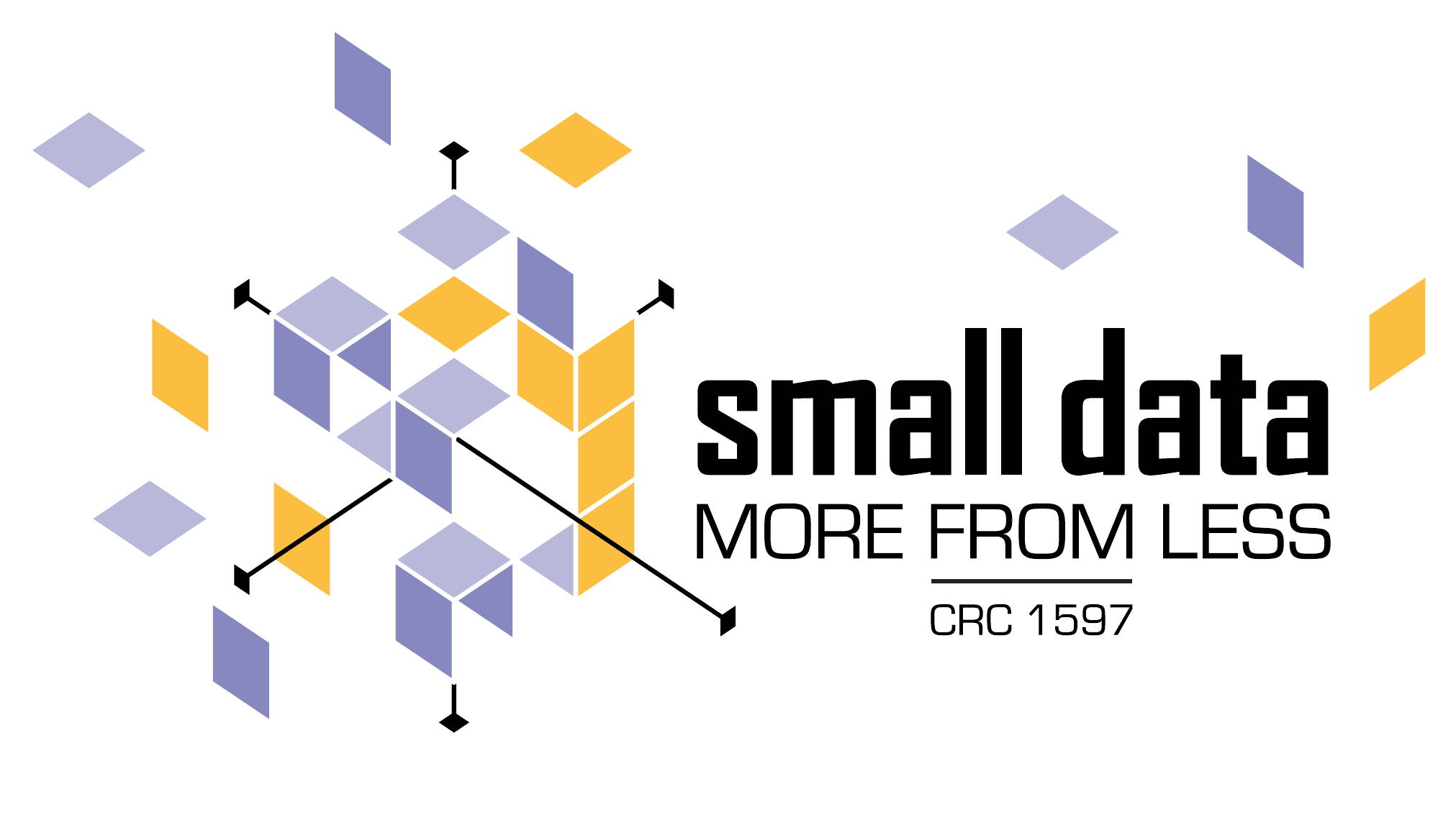}
        };
    \end{tikzpicture}
}
\makeatother

\usepackage{tcolorbox}
\tcbuselibrary{breakable}
\usepackage[labelfont=bf]{caption}

\definecolor{similaritycolor}{HTML}{b8c7d9}
\definecolor{boxcolour}{HTML}{e8e8e8}

\usepackage{newfloat}
\DeclareFloatingEnvironment[fileext=frm,placement={!ht},name=Box]{myfloat}

\makeatletter
\renewcommand{\maketitle}{
    \begin{center}
        \addlogo
    \end{center}
    \vspace{1cm}
    \noindent\begin{minipage}{\textwidth}
        \justifying 
        {\fontsize{22}{24}\selectfont \bfseries \@title \par} 
        \vspace{1cm}
        {\justifying \@author \par}
    \end{minipage}
}
\makeatother

\title{{Small Data Explainer -- The impact of small data methods in everyday life}}

\author[1,2,10]{Maren Hackenberg\,\raisebox{0.1em}{\footnotesize\faEnvelope}}
\author[1]{Sophia G. Connor}
\author[1]{Fabian Kabus}
\author[3]{June Brawner}
\author[3,4]{Ella Markham}
\author[3]{Mahi Hardalupas}
\author[3]{Areeq Chowdhury}
\author[5,6]{Rolf Backofen}
\author[7]{Anna Köttgen}
\author[8]{Angelika Rohde}
\author[2,9]{Nadine Binder}
\author[1,2,6]{Harald Binder}
\author[ ]{the Collaborative Research Center 1597 Small Data}

\affil[1]{Institute of Medical Biometry and Statistics, Faculty of Medicine and Medical Center, University of Freiburg, Freiburg, Germany}
\affil[2]{Freiburg Center for Data Analysis, Modeling and AI, University of Freiburg, Freiburg, Germany}
\affil[3]{The Royal Society, London, United Kingdom}
\affil[4]{University of Edinburgh, Edinburgh, United Kingdom}
\affil[5]{Department of Computer Science, Faculty of Engineering, University of Freiburg, Germany}
\affil[6]{Centre for Integrative Biological Signalling Studies (CIBSS), University of Freiburg, Freiburg, Germany}
\affil[7]{Institute of Genetic Epidemiology, Faculty of Medicine and Medical Center, University of Freiburg, Germany}
\affil[8]{Department of Mathematical Stochastics, Faculty of Mathematics and Physics, University of Freiburg, Germany}
\affil[9]{Institute of General Practice/Family Medicine, Faculty of Medicine and Medical Center, University of Freiburg, Germany}
\affil[10]{Present address: European Molecular Biology Laboratory, Heidelberg, Germany}

\date{}

\begin{document}

\maketitle

\small \noindent\faEnvelope~Corresponding author:  
\href{mailto:maren.hackenberg@embl.de}{maren.hackenberg@embl.de}
\normalsize

\begin{abstract}
{\setlength{\leftskip}{0.5cm}
\setlength{\rightskip}{0.5cm}
The emergence of breakthrough artificial intelligence (AI) techniques has led to a renewed focus on how small data settings, i.e., settings with limited information, can benefit from such developments. This includes societal issues such as how best to include under-represented groups in data-driven policy and decision making, or the health benefits of assistive technologies.
We provide a conceptual overview, clarify the relationship between small data and big data, and identify common themes from exemplary case studies and application areas. Potential solutions are described in a more detailed technical overview of current data analysis and modelling techniques, highlighting contributions from different disciplines, such as knowledge-driven modelling from statistics and data-driven modelling from computer science. By linking application settings, conceptual contributions and specific techniques, we highlight what is already feasible and suggest what an agenda for fully leveraging small data might look like.
\par}
\end{abstract}
\clearpage

\section{Introduction}\label{sec:introduction}

In recent years, there has been an increasing recognition of the potential of small data in technology development and decision-making processes, particularly as a complement to big data approaches \cite{faraway2018when, centerforsecurityandemergingtechnology2021small, ehrenberg-silies2019big, strickland2022andrew, rather2024breaking}. 
This includes renewed interest in methods for extracting knowledge from limited information, which have long been central in statistics, applied mathematics, and scientific inquiry.
While there is no consensus on the definition of small data, small data approaches are typically needed in scenarios where there is a limited amount of information, such as in the analysis of a small number of observations in a dataset. However, limited information is not determined purely by sample size. Whether a dataset is effectively small also depends on the amount of heterogeneity between observations. For example, a clinical patient dataset with six patients might be considered small, while an experiment with six isogenic mice in biomedical research would not be considered small, due to large diversity in human beings as compared to animal models in biomedical research. A further important factor is the complexity of the questions to be explored with a dataset. For example, training large language models (LLMs) with machine learning techniques requires vast training datasets of millions or billions of documents. The training of a specialised LLM with only thousands of documents then is a small data challenge, while the use of simpler statistical modelling techniques with clinical patient data might only face a small data challenge when the number of observations is below a few dozen patients.

Big data approaches can provide powerful tools when the available data adequately cover the population, its heterogeneity, and the relevant subgroups and decision contexts of interest. However, there are many scenarios where large datasets are not available or where unique and specific data points can provide critical understanding \cite{ferguson2014big, faraway2018when}. Instead of relying predominantly on data volume to identify overall trends and patterns, small data methodology can provide valuable insights from smaller and often more homogeneous datasets, which might put the focus on specific, sometimes overlooked subgroups of a population and thus address diversity in a targeted way \cite{faraway2018when, hekler2019why}. 
Even when there are large datasets available, they may exclude under-represented groups, or fail to capture rare combinations of characteristics that matter for a specific decision, or small groups might not be represented in patterns extracted by big data techniques, reinforcing biases \cite{shilo2020axes, leonelli2019data}.
In other settings, collecting more data may be impossible, unethical, too costly, or misaligned with data minimisation principles. 
Small data approaches, on the other hand, make these limitations explicit and provide techniques for reasoning with scarce, context-specific information \cite{hekler2019why, faraway2018when}. Therefore, small data methods are seen as a fruitful and necessary part of the methodological toolkit \cite{rather2024breaking}.

Classical statistical methodology, including experimental design, hierarchical modeling, regularisation, and uncertainty quantification, has a long-standing tradition in addressing many of these challenges, and remains foundational to modern small data approaches \cite{faraway2018when, faes2022artificial, vandeschoot2021bayesian}.
As small data problems occur in a range of fields, small data methodologies have also been developed across many different research areas (e.g., computer science, mathematics, statistics) \cite{li2020learning, shaikhina2017handling, dou2023machine}.  While this shows the relevance of small data research in many domains, it also means that research may be impeded due to a lack of interdisciplinary communication \cite{kokol2022machine, centerforsecurityandemergingtechnology2021small}.  In particular, the lack of a shared language for small data approaches across disciplines is a major hurdle, in addition to technical challenges when using small data methods, such as overfitting and validation \cite{faes2022artificial, bornschein2020small}. To foster such a shared language, this explainer jointly addresses audiences from many different communities developing small data methods, aiming to situate recent AI- and technology-oriented perspectives on small data within a broader methodological landscape.

Specifically, we aim to raise awareness of, and illustrate, the unique challenges posed by small data, which originate from limited sample sizes and limited information in the face of complex data analysis tasks. This is relevant to a range of audiences, including: 1) policymakers and policy professionals who are interested in understanding how advancements in data science can inform policy decisions; 2) data scientists familiar with statistical methods, modelling techniques, and machine learning algorithms, who are interested in understanding small data techniques, challenges, and potential solutions; and 3) computer scientists, mathematicians, and statisticians developing algorithms, tools, and techniques for processing and analysing data.

Beyond reviewing existing approaches, we aim to contribute a policy-oriented perspective on how small data research priorities are set. We argue that policymakers influence not only the adoption but also the direction of small data methodologies, by defining application domains in which small data challenges are most consequential. Recognising this relationship is essential for aligning methodological innovation with evidence-informed policy needs.

This explainer was guided by three primary questions, answered through extensive desk review and evidence synthesis:
\begin{enumerate}
    \item What are the fundamental principles underlying the concept of small data, and how do they differ from those of big data, particularly in the context of technology development and decision-making?
    \item Why has there been a growing interest in small data approaches, and what are their implications for inclusive technology development and decision-making processes?  
    \item How can various methods and techniques for working with small data be adapted and optimised to address the unique challenges and opportunities present in the context of AI development and personalised applications?
\end{enumerate}

To approach these questions, we look at small data from multiple perspectives. Section \ref{sec:smalldatapolicy} focuses on the relevance of both big and small data for policy and decision making, and the core components of small data applications. We first discuss how seemingly large datasets may nevertheless become small data challenges when the relevant information for a specific subgroup, context, or decision is sparse. This includes settings where under-represented and minority groups, rare combinations of characteristics, or context-specific needs may be insufficiently represented in the evidence used for decision-making, which is especially concerning for high-stakes policy decisions. This section also includes case study examples of small data use (e.g., rare disease treatment and wearable assistive technology), and introduces the themes of similarity, transfer, and uncertainty for structuring small data applications and the subsequent discussion of specific methods. It is targeted towards a broad audience with a specific focus on policymakers.

In Section \ref{sec:exemplaryapplications}, also targeted towards a broad audience, we highlight some exemplary application fields of small data. For some of these, some tailored small data approaches have already been developed, such as for rare disease settings or precision medicine, which could serve as templates for other application fields, where small data approaches are still lacking. 

In Section \ref{sec:methodsworkingsmalldata}, we describe small data methodologies in more detail (e.g., approaches for incorporating similarity, foundation models, few-shot learning, meta learning, and the combination of knowledge-driven and data-driven modelling). Many of these methodologies were developed by different disciplines and fields, which may potentially hinder awareness and adoption of methods by different research groups. To address this, we highlight the contributions of different disciplines and explore potential links. We also present some common challenges in using small data, such as overfitting and validation. This section therefore takes a more technical approach and is tailored for an interdisciplinary data science audience. 

We conclude with some potential opportunities and propose an agenda for increasing the use of small data in Section \ref{sec:conclusionsperspectives}.

\section{Small data and policy}\label{sec:smalldatapolicy}
 
\subsection{The significance of big data for research and decision making}\label{subsec:sig_big_data_research}

The ability to process vast amounts of information quickly and efficiently has made possible the identification of broad trends, prediction of future behaviours, and optimisation of complex systems. These discoveries and innovations are fuelled by so-called big data. The term big data initially referred to datasets so vast they required specialised computing solutions for analysis \cite{manovich2012trending}, yet such advanced computing power nowadays is easily available. Despite the name, sheer size is not what defines big data. In simple terms, big data is more accurately characterised by the ability to search, aggregate, and cross-reference extensive datasets, with potential challenges due to a large variety of data types and data quality issues \cite{shilo2020axes}. From a sociotechnical lens, big data rests at the intersection of 1) technology (maximising the power of computation and algorithms to gather and analyse large datasets); 2) analysis (identifying patterns within large datasets to make claims about processes); and 3) wider narratives about the capacity of large datasets to provide objectivity or accuracy previously not possible \cite{boyd2012critical}.

The contemporary emphasis on big data began in the early 2000s, coinciding with advancements in computing power and the large varied quantities of data generated and collected by digital technologies, including those connected through the internet. Global technology companies began to harness big data to transform their business models, creating highly personalised user experiences \cite{mayer-schonberger2013big, davenport2013big}. The success of these enterprises underscored the value of big data, leading to the widespread adoption of big data practices across various sectors, including healthcare, finance and public policy \cite{hillyer2020heres}. 

The corresponding increased awareness of the opportunities of data and focus on gathering data for better understanding structures and processes has led to significant advancements, regardless of whether specific applications are big data or not. For example, data availability in healthcare has enabled breakthroughs in precision medicine \cite{ashley2016towards} and predictive maintenance in manufacturing \cite{lee2013recent}. More than this, big data created ``a radical shift in how we think about research […] Big Data reframes key questions about the constitution of knowledge, the processes of research, how we should engage with information, and the nature and the categorization of reality''\cite{boyd2012critical}.

\begin{myfloat}[tb]
\begin{center}
    \begin{tcolorbox}[colback=boxcolour, colframe=boxcolour,  width=1.0\textwidth, boxrule=0mm, sharp corners,
    before=\vspace{-2em},after=\vspace{-0.5em}]
         A clinician specialising in Gaucher disease, a rare genetic disorder that leads to the accumulation of glucocerebroside in various organs \cite{stirnemann2017review} due to the deficiency of a specific enzyme, wants to assess whether a patient is responding well to enzyme replacement therapy (ERT). It is often challenging for the clinician to find the correct dose of ERT, as the disease presents with diverse manifestations and many different factors need to be considered \cite{andersson2005individualization}. In order to address these issues, the clinician might have access to a clinical registry of patients with this medical condition, which contains information as to their ERT doses. 
         However, because Gaucher disease is rare, the registry is likely to be small. Using a machine learning approach for dose prediction can lead to a small data challenge, as the available data may not be enough to directly guide dosing decisions.

        When the clinician has a new patient, for example a 6-year-old  child trialling ERT again after a pause, the machine learning approach should take into account how similar the new patient is to the patients already present within the clinical registry. Often, there will be considerable heterogeneity across patients in a registry, which means that patients are dissimilar to each other. Therefore, the new patient might need to be matched to the most similar subgroup to allow for modelling and prediction based on this subgroup, while still taking some information from the other subgroups into account. In this case, for example, there might only be a limited subgroup of patients under the age of 10. For these reasons, it can be difficult to match the current patient to previous patients in order to make accurate predictions. When the new patient cannot be appropriately matched, transfer techniques are required to still enable prediction modelling by at least using some amount of information, e.g., by extrapolating across age groups.
    \end{tcolorbox}
    \end{center}
    \vspace{-0.5em} 
    \caption{Hypothetical exemplary scenario from the field of rare disease small data applications.} 
    \label{box:raredisease} 
\end{myfloat}

\subsection{Seemingly large datasets as small data settings}\label{subsec:seemingly_big_data_limits} 

A dataset may be large overall, while still containing limited relevant information for the specific question or subgroup at hand. In such situations, the effective amount of usable information can be small, even if the aggregate dataset would usually be described as big data. This is particularly important in high-dimensional settings, i.e., in settings with a large number of relevant characteristics and possible combinations of them, so that even ``big'' datasets may contain few observations for the particular case under consideration. 

If the available data adequately represent the population and decision context of interest, big data approaches can provide powerful tools for identifying patterns, capturing variation, and supporting personalised decisions. 
However, in some contexts, large datasets are simply not available, especially in specialised fields like rare diseases or niche markets (see Box \ref{box:raredisease} for an exemplary scenario, and Section \ref{subsec:rarediseases} for more details). 
A registry may be the largest available dataset for a disease and may still be sparse for a particular clinical decision, such as dosing for a young child with a specific treatment history. 
Similar issues arise when designing technologies or policies for under-represented groups. Additional data collection may be infeasible or unethical. Even where it is possible, extensive additional collection may fail to capture the subgroup or situation that is most important for the decision to be made, as relevant subgroups may be difficult to identify in advance or represented only sparsely.

In these cases, relying on data volume alone can create a false sense of evidential security. Then, small data approaches can be useful because they focus explicitly on detailed, context-specific information from scarce data.
In our view, the distinction between small data and big data methods is therefore not defined purely by dataset size. Rather, small data methods represent a complementary mindset and awareness of how to approach a dataset, as they draw attention to how much information is actually available for the target task or decision at hand, to what extent information can be transferred from related settings or other sources of knowledge, and how to quantify and incorporate uncertainty.

\begin{figure}[t]
    \centering
    \includegraphics[width=1\textwidth]{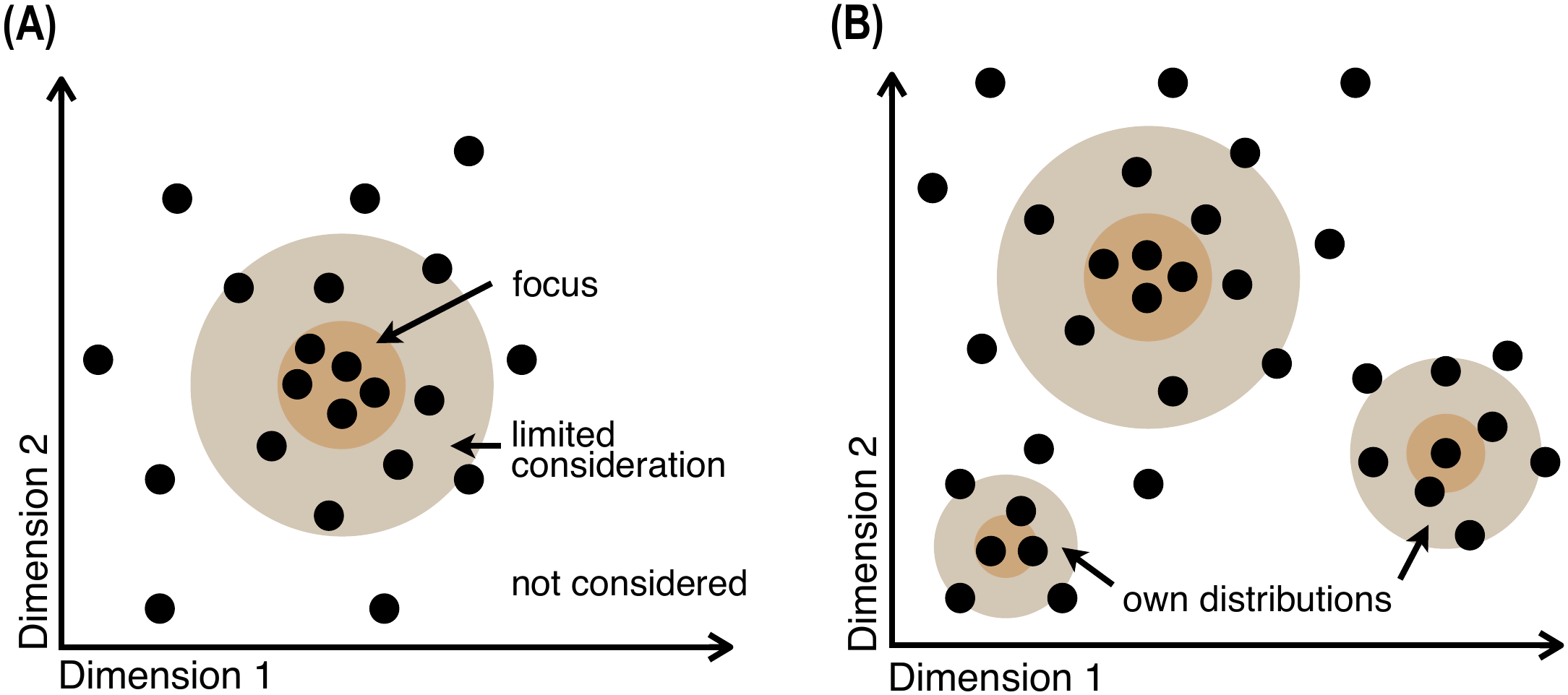}
    \caption{Conceptual illustration of small data challenges with hypothetical data where individuals are described by two dimensions, such as ``age'' and ``years of experience with digital technologies'' when designing a chatbot system based on large language models (LLMs). (A) Small data conceptualised as limited information about individuals at the margins of a distribution or about rare combinations of characteristics. The individuals in the ``Focus'' area are within the average, whereas individuals that deviate from this average are considered less often or not at all. (B) Conceptualisation of small data as subgroups of individuals which might be best described by their own distribution, e.g., not treating children as small adults, or analysing the specific needs of elderly users.}
    \label{fig:smalldatachallenges}
\end{figure}

A historical concern with averages helps explain why this distinction is especially relevant for policy. In the 19th century, Belgian astronomer and statistician Adolphe Quetelet FRS introduced the concept of the ``average man'' (``l'homme moyen''), a hypothetical individual representing the average values of human traits in a population \cite{tafreshi2022adolphe}. By applying the Gaussian distribution, which had been used primarily in astronomy and other physical sciences, to human traits, he suggested that most people cluster around a central, average value, with fewer individuals exhibiting extreme values at the margins. 
Quetelet's work illustrates the historical importance of population averages as summaries for reasoning about human populations in statistics and social science \cite{tafreshi2022adolphe}. Although contemporary big data methods have moved much beyond this framework, the example remains relevant for policy because population-level summaries can still shape how problems, target groups, and interventions are defined.
Such a focus on averages can provide useful summaries, but it can also hide variation that is central for a particular decision. This is an issue not so much of raw data volume, but of the focus when looking at the data. Instead of regarding individuals with less common traits or needs as the extreme values of a distribution (Figure \ref{fig:smalldatachallenges}, Panel A), a small data-oriented mindset offers an alternative conceptualisation, where small groups may need to be considered as having their own distribution (Figure \ref{fig:smalldatachallenges}, Panel B). This implies the need to reason about these distributions and subgroup structures, which requires methodology for similarity assessment, transfer of information from other subgroups, and uncertainty quantification.

In policy decisions, marginalisation of individuals and scenarios that do not fit a majority pattern can lead to knowledge gaps and inadequately addressed needs, and to ineffective or inequitable decision-making. Instead, small data-aware decision-making can foster more inclusive and effective policy and technological solutions, ultimately benefiting the entire population.

\subsection{The rise of small data}\label{subsec:rise_small_data}

Small data is increasingly relevant because many contemporary data-driven applications require decisions in contexts where the information available for the specific task, subgroup, or individual remains limited. This includes settings where data are scarce in an absolute sense, as well as settings where larger datasets need to be adapted to a more specific context.

\begin{myfloat}[t]
\begin{center}
    \label{box:foundationmodels} 
    \begin{tcolorbox}[colback=similaritycolor, colframe=lightgray, sharp corners, width=1.0\textwidth, boxrule=0mm]
    \textbf{Glossary: \textit{Foundation models}} are large-scale machine learning models trained on vast amounts of diverse data, allowing them to serve as a general-purpose base for a wide range of downstream tasks. These models, such as large language models (LLMs) and multimodal AI systems, acquire broad representations that can be adapted to more specific applications with limited available data through fine-tuning or in-context learning. These capabilities make them a powerful tool for both big and small data settings.
    \end{tcolorbox}
    \vspace{-1em} 
\end{center}
\end{myfloat}

Small data is sometimes referred to as the ‘original data’: early-modern scientific discoveries, from those of Darwin's evolutionary theory to Einstein's fundamental laws, predate big data approaches and contemporary computational abilities. Humans are also naturally small data learners: given a single image of a car, children can generalise the concept and recognise similar objects. This is an impressive example of one-shot learning, i.e., learning a class from a single labelled example, which is not straightforward to obtain with classical machine learning approaches \cite{vinyals2016matching}. Therefore, it might be attractive to use small data ideas also in big data contexts. Certain methodologies already have integrated small data with big data to transfer information between the big dataset and the small dataset. For example, foundation models now allow the transfer of information from a huge body of text, images, or other data to settings with limited data \cite{bommasani2022opportunities} (see Box \ref{box:genAI} for an exemplary scenario).

\begin{myfloat}[!tb]
\begin{center}
    \begin{tcolorbox}[colback=boxcolour, colframe=boxcolour,  width=1.0\textwidth, boxrule=0mm, sharp corners,
    before=\vspace{-2em},after=\vspace{-0.5em}]
         Suppose a policymaker in a large tech company aims to improve the security and robustness of the company’s software products against cybersecurity threats, e.g., in accordance with a report recently published by the White House Office of the National Cyber Director \cite{nationalcyber2024back}. To achieve this, they focus on addressing known classes of vulnerabilities through use of secure software development practices. The most commonly used programming language for the company’s products is C++, which is also one of the languages that lacks memory safety \cite{departmentofhomelandsecurity2023case}. In addition, the policymaker knows that it is common practice among developers in their company to use LLM-based products for assistance when generating code. While recent versions of C++ allow for memory-safe programming, the C++ code used to train LLMs is likely to span several decades, i.e., represent old non-safe practices with potential memory safety bugs \cite{khoury2023how}. Thus, the challenge would be to use limited amounts of memory-safe examples of code to guide the LLM in generating further memory-safe code for the requested task. Ideally, there would be a similarity metric to assess different LLMs and determine which can best provide this code, in order to select the best solution. Alternatively, LLMs could be fine-tuned to generate safer code. Users should at least be provided with a warning if the generated code deviates from the safety standards of the company, or the model might even be able to more precisely indicate how confident it is that its output is indeed memory-safe.
    \end{tcolorbox}
    \end{center}
    \vspace{-0.5em} 
    \caption{Hypothetical exemplary scenario from the field of generative AI and foundation models. }
    \label{box:genAI} 
\end{myfloat}

Consequently, small data is (re-)emerging as a crucial complement to big data, in part because of its relevance to advances in technologies like AI. This approach is particularly important in areas where context-specific insights are critical, such as in rare disease research (see Box \ref{box:raredisease}), and personalised marketing strategies (see Box \ref{box:trustedpersonalisedrecommendations}). 

\newpage 

\begin{myfloat}[h]
\begin{center}
    \begin{tcolorbox}[colback=boxcolour, colframe=boxcolour,  width=1.0\textwidth, boxrule=0mm, sharp corners,
    before=\vspace{-2em},after=\vspace{-0.5em}]
         Suppose a finance tech start up (``FinTech'') is developing an app that helps its users make informed financial decisions by providing them personalised recommendations. As consumer trust is an important factor for the adoption of the product \cite{stewart2018data, torno2021robowhat}, they focus on a strong adherence to user privacy, in particular, by storing information only on the device of the user, e.g., a smartphone. In addition, the recommendation algorithm is focused on data minimisation, by asking the user only for absolutely necessary information and storing only aggregate data where possible, e.g., only the average proportion of income spent on rent or eating out. This limits the privacy impact, e.g., in case of a data breach. 

        To enable precise predictions as a basis for recommendations, missing information that is still required might be extrapolated from similar users, e.g., from a database of users who agreed to such use of their data. For example, the prediction algorithm might determine how the individual fits different archetypes, i.e., a student who likes to travel or a married individual with a stable income and children, using similarity measures. As banks continue to open up their data to third parties \cite{wewege2020disruptions}, the app should give specific recommendations on retirement planning by transferring the generic recommendation model onto the specific spending habits of the user provided by the bank. As financial advice, especially concerning investments, involves uncertainty, which might be increased by data minimisation, the app should include
        uncertainty evaluations to help the individuals make informed decisions about the risks involved, e.g., when deciding for a saving strategy. The app might even indicate what additional information would be needed to reduce uncertainty.
    \end{tcolorbox}
    \end{center}
    \vspace{-0.5em} 
    \caption{Hypothetical exemplary scenario from the field of finances and trusted personalised recommendations.}
    \label{box:trustedpersonalisedrecommendations} 
\end{myfloat}

The growing interest in small data is also driven by advancements in data collection and analysis technologies that make it easier to gather and interpret more detailed information from smaller samples. For example, wearable devices, mobile sensors, and specialised software enable the continuous collection of granular data, data that is very detailed and specific, providing a rich source of insights that were previously inaccessible \cite{piwek2016rise} (see Box \ref{box:assistivetechnologies}). 

As another example, mixed-methods research combines quantitative data from large samples with qualitative insights from smaller groups to provide a more comprehensive understanding of complex phenomena. This integrative approach is particularly valuable in fields like social sciences, where understanding human behaviour requires analysis of both broad patterns and individual experiences. For a primer on these themes see \cite{tashakkori2003handbook}. Small data techniques are also relevant where limited, granular datasets are collected by individual researchers during day-to-day experiments, as these datasets may be richly diverse, covering a range of heterogeneous scenarios, and relatively small, so-called long-tail data \cite{ferguson2014big}. Long-tail data is valued for its diversity and specificity, which can provide critical insights that large, aggregated datasets might overlook. 

Furthermore, small data approaches can help foster innovation and resilience in technologies, as they specifically acknowledge subgroups, potentially in a wide range of scenarios, making them more robust in the face of dynamic environments. Thus, personalised and context-specific technology solutions will rely on the use and inclusion of small data.

\begin{myfloat}[!tb]
\begin{center}
    \begin{tcolorbox}[colback=boxcolour, colframe=boxcolour,  width=1.0\textwidth, boxrule=0mm, sharp corners,
    before=\vspace{-2em},after=\vspace{-0.5em}]
         Suppose a healthcare provider aims to decrease the number of hospital admissions resulting from recurrent falls among patients. To do this, they give wearable devices with accelerometers to individuals with a prior history of falls \cite{ramachandran2020survey}. The detection of falls must take place on the device with the limited data available there. After a period of time, there is enough accelerometer data available for the individual, such as how they normally move, so that personalisation and thus improved fall detection might be feasible. If the provider has a database of accelerometer data generated from use of this wearable device, the data from the individual could be mapped to subgroups of other individuals, or alternatively, a prediction model could be used that is aware of subgroups and has already been trained on the database. Ideally, this analysis in the context of similar data should be performed without transferring the data of individuals to a central server, to protect privacy. 

        However, fall detection using accelerometer data is challenging due to significant variability in how movements are recorded, as well as differences in individual anatomy \cite{igual2013challenges, medrano2016effect}. This variability might introduce considerable uncertainty in similarity assessment and predictions. Therefore, approaches for quantifying uncertainty in all steps are needed. In addition, some of the patients might not like the wearable device, but want to use the accelerometer on their phones instead \cite{martinez-villasenor2019upfall}. This means that information transfer is required to link this new data type to the original datasets and models.
    \end{tcolorbox}
    \end{center}
    \vspace{-0.5em} 
    \caption{Hypothetical exemplary scenario from the field of assistive technologies.} 
    \label{box:assistivetechnologies} 
\end{myfloat}

\subsection{Common themes: similarity, transfer, and uncertainty}\label{subsec:commonthemes}

\begin{figure}[!tb]
    \centering
    \includegraphics[width=1\textwidth]{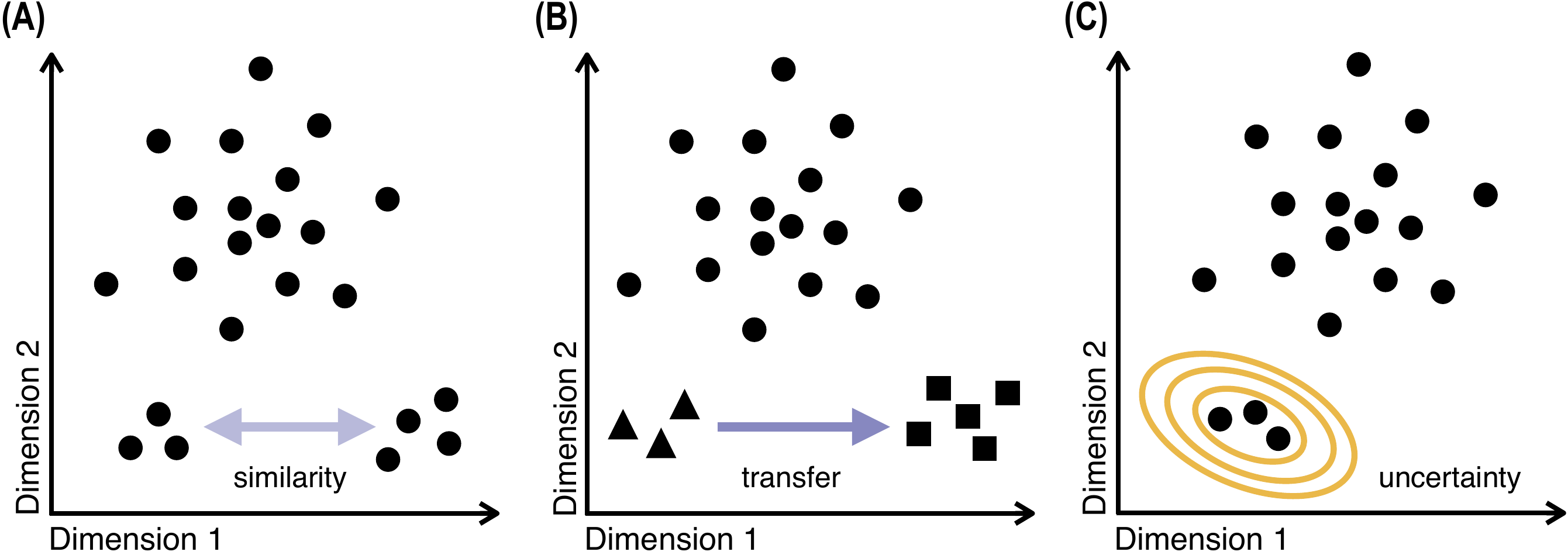}
    \caption{Illustration of common themes, namely similarity, transfer, and uncertainty, when addressing small data challenges. The example uses hypothetical data where individuals are described by two dimensions, such as when considering factors ``age'' and ``years of experience with digital technologies'' when designing a chatbot system based on LLMs. (A) Assessing the similarity of groups for potentially incorporating information, e.g., deciding whether to jointly analyse data from children and older adults when considering inclusive chatbot design. (B) More general information transfer, when data/information cannot be directly combined, e.g., leveraging child-centered user experience described in the scientific literature when designing a chatbot for older adults. (C) Quantifying uncertainty, to assess whether there is enough data for obtaining reliable evidence, e.g., deciding whether there is enough information to reasonably tailor a chatbot for children.}
    \label{fig:commonthemes}
\end{figure}

For making small data challenges manageable, it can be useful to identify sub-tasks that appear in many small data settings. In the examples given above (in particular Boxes 1-4), there were three themes that repeatedly emerged: similarity, transfer, and uncertainty. These are illustrated in Figure \ref{fig:commonthemes}, and are briefly introduced in the following section. A detailed description of the corresponding technical approaches is provided in Section \ref{sec:methodsworkingsmalldata}.

\subsubsection{Similarity}\label{subsec:similarity}
Similarity is relevant when comparing different datasets, containing data of individuals, for potential subsequent integration. Integrating datasets increases the number of observations available for modelling, and involves assessing the similarity of individuals to other individuals or groups of individuals, such as in rare disease applications, to leverage evidence from similar individuals. In other settings, similarity of datasets that include several observational units needs to be assessed, e.g., when deciding whether to combine datasets for modelling. As similarity will typically be assessed quantitatively, modelling techniques that are subsequently used also need to incorporate such quantitative information on similarity.

When there is already a big dataset or (pre-)trained model, e.g., a foundation model, approaches for similarity can be useful for assessing whether a dataset at hand is close enough to the typical settings represented by the big dataset or model to warrant use. Subsequent adaptation, when there is sufficient similarity, might then include personalisation of the model, e.g., by identifying subgroups represented by the large model that are the most similar to the dataset at hand. This corresponds to acknowledging the heterogeneity of big datasets and potentially reducing them into interlinked small datasets.

\subsubsection{Transfer}\label{subsec:transfer}
The combination of similar datasets already implicitly entails a transfer of information from one dataset to the other. Yet, we understand transfer in a broader sense, relating not only to the transfer of information between similar datasets, but also to a more general information transfer from external sources to enrich a small dataset. In particular, information from other sources to be incorporated can include, e.g., other types of data (other data forms, databases, etc., see Figure \ref{fig:commonthemes}, Panel B) or pre-trained models (e.g., LLMs), reflecting implicit knowledge, to facilitate modelling with small data by additional prior knowledge. This requires methods for explicitly transferring information. Use of methods for transfer ideally should include some quantification of similarity between the different types of information sources to assess whether information transfer indeed is beneficial, or whether it instead is detrimental, e.g., when LLMs override the information in the dataset at hand that is used as context.

In some settings, knowledge of the domain at hand might not be available in formats ready for information transfer, such as mechanistic knowledge or heuristic knowledge of experts. A formalisation step is then required, before information transfer can be performed. For example, knowledge on relations between quantities of interest could be expressed in the form of mathematical equations, even if it is not possible to fully and precisely specify these equations. Methods for information transfer then need to incorporate the uncertainty in specification.

\subsubsection{Uncertainty}\label{subsec:uncertainty}
Due to limited information, uncertainty is a particularly pressing issue to be considered in small data settings. While uncertainty has been conventionally considered in the context of variability in parameter estimation in statistical models, such as when estimating treatment effects, the term can also be understood in a broader sense. In particular, there are many other kinds of uncertainties, including general sources of uncertainty in modelling, such as the choice of appropriate models \cite{hullermeier2021aleatoric}. Some sources of uncertainty are more specific to small data, particularly uncertainty arising from quantification of similarity and transfer of information, e.g., in the formalisation of expert knowledge for inclusion into modelling.

Where uncertainty cannot be avoided, comprehensive quantification of the different sources of uncertainty is needed. This enables informed use of the resulting models and predictions. Such quantification can also enable deliberate acceptance of uncertainty, e.g., when making informed decisions on data minimisation for a trade-off between uncertainty and privacy (see Box \ref{box:trustedpersonalisedrecommendations} for an example).

\subsection{Implications of small data for evidence-informed policy-making}\label{sec:implicationspolicymaking}
The policy implications of small data approaches are significant. During times of financial constraint, decision-makers often rely on data-supported strategies to set spending priorities and justify investments in public funds. However, this reliance can marginalise individuals who do not fit within the majority, such as those with disabilities or exceptional access requirements \cite{nationalresearchcouncilus2011precision}.

For evidence-informed policymaking, the central question is therefore not only how much data is available overall, but whether the available evidence is informative for the people, contexts, and decisions at stake. Data-driven decision-making can rely on population-level summaries that are appropriate for many purposes, but may be insufficient for groups or situations that are sparsely represented in the underlying data. This is particularly relevant for assistive technologies, where evidence is often drawn from small and heterogeneous samples, and where the needs of users may differ substantially across contexts \cite{ciardiello2014delivering}. Small data approaches can help make these limitations explicit by focusing attention on subgroup representation, transfer of relevant information, and uncertainty.

Focusing on under-represented groups can also lead to better outcomes for both individuals and society as a whole. 
An example where approaches developed for a minority have already come to benefit the larger population is closed captioning \cite{downey2008closed}. Initially developed for the deaf and hard of hearing, it is now widely used in noisy environments like gyms and bars, and by individuals learning new languages or trying to follow along without sound. Similarly, automatic doors \cite{steinfeld2012universal}, Text-to-speech (TTS) and Speech-to-text (STT) systems \cite{harper2019web}, voice assistants \cite{pradhan2018accessibility} like Amazon's Alexa and Apple's Siri, as well as smartphone accessibility features \cite{ellis2015disability} were also originally designed to assist disabled people but have developed a much wider scope given their convenience and ease of use in various situations. 

Beyond recognising the importance of small data, policymakers play a critical role in shaping the direction of methodological research. 
The recent advances in small data methodologies demonstrate that meaningful modelling and inference are increasingly possible even for small or under-represented population groups. However, the suitability, development and maturation of such methods are strongly shaped by the application contexts in which they are expected to operate. 
As a result, their effective use in evidence-informed policymaking requires prioritisation and agenda setting beyond technical innovation alone.
Here, policymakers and funding bodies can play an active and decisive role in shaping the direction of methodological innovation, by explicitly articulating priority domains, such as specific health, social, or assistive technology contexts where small data challenges are most consequential. By signalling where robust small data solutions are most urgently needed, policymakers can help align research incentives and funding priorities with societal needs. Such priority setting fosters development of tailored methods and can help ensure that small data methods evolve in ways that directly support evidence-informed policy making.

\section{Exemplary application fields of small data}\label{sec:exemplaryapplications}

\subsection{Rare diseases and N-of-1 studies}\label{subsec:rarediseases}
Small data methods are crucial for studying rare diseases, i.e., diseases afflicting less than one in 2,000 persons \cite{villalon-garcia2020precision}, as this is a prime example for medical conditions with limited data (see Box \ref{box:raredisease} for a hypothetical example). Correspondingly, this area has produced numerous methodologies specifically designed for small datasets. Still, it is a challenge to adapt machine learning, in particular deep learning techniques, which typically require more data \cite{banerjee2023machine}. Nevertheless, such methodological developments can be used as a blueprint for other small data settings. It is important to note that many of these techniques are not explicitly labelled as small data methods, even though they apply to these scenarios, reflecting the challenge of the lack of shared terminology around small data methods.

N-of-1 studies are a particular type of small data study which produce data limited by design, as they focus on collecting and analysing data from a single participant \cite{mcdonald2017state}. This is often performed in order to determine how effectively the participant responds to a specific intervention, and may involve combining multiple different types of data from the individual, which requires transfer methodologies. N-of-1 studies are better suited for chronic conditions rather than rapidly progressive diseases, and commonly deployed to assess neurological or musculoskeletal disease \cite{mirza2017history}. This is because chronic conditions are often relatively stable or slowly progressive, and require long term management, which enables multiple crossover periods to assess efficacy of various treatments in the same patient.

Small data methods can assess medication efficacy, as well as the efficacy from surgical procedures, behavioural modifications, or medical device assessments \cite{gabler2011nof1}. These methodologies can then be applied to similar study designs where only a single participant is assessed. When aggregating many N-of-1 studies, the combined data could, e.g., be used for future predictions \cite{zucker2010individual}. Yet, to do this, one must assess how similar the datasets are and develop methods to combine datasets when they are not in exactly the same format, i.e., tailored approaches are needed for information transfer.

\subsection{Precision medicine and case-based reasoning}\label{subsec:casedbasedreasoning}
Personalisation of models, predictions and decisions in small data settings aims to tailor particular outcomes to the unique characteristics of a single observational unit or a small group. In precision or personalised medicine, this involves adapting specific therapies to suit each individual, accurately diagnosing diseases, or predicting individual health outcomes by taking into account unique genetic, lifestyle or environmental factors \cite{nationalresearchcouncilus2011precision}. Effectively, this means that even a rather large, potentially multi-modal dataset might be reduced to many small subsets when looking for homogeneous groups of patients that could receive a personalised prediction. Therefore, much like in rare disease settings, many methodologies developed in the field of precision medicine are useful for small data settings, though they are often not explicitly identified as such. Often, methodologies initially developed for specific diseases are also applicable to other small data scenarios once researchers recognise similar underlying features.

Oncology has consistently led to advancements in precision medicine, largely because of the genetic basis of many cancers and the rapid progression of genomic technologies, e.g., allowing for more reliable and affordable cancer diagnosis \cite{ciardiello2014delivering, schmidt2016precision}. These advances enable the comprehensive screening of an individual’s genome in just days, facilitating the development of tailored diagnostic, prognostic and therapeutic strategies. Large-scale analyses are effective at identifying patterns across large datasets, but small data methods are more tailored to tasks such as detecting rare mutations, or analysing data from an individual’s own cells to predict how they will respond to treatment through personalised in vivo and in vitro cancer models \cite{walker2019evidence}. However, this raises many small data challenges, such as looking for similar data to increase the dataset size, transferring information from other sources, and assessing the uncertainty, i.e., quantifying the reliability of predictions from such cancer models.

The paradigm of case-based reasoning (CBR) considers an extremely personalised setting where the data from the patient at hand, i.e., a single observational unit, is used as a starting point for reasoning. CBR then offers a modelling methodology for using previous similar examples of a problem and its solutions, each called a ``case'', to offer recommendations for a current similar issue \cite{aamodt1994casebased}. This recommendation is then stored for use with future cases, allowing for iterative learning \cite{dussart2008optimizing}. As for other settings in precision medicine, similarity assessment is therefore crucial in CBR because the accuracy of the solution depends on the accuracy of the evaluation of similarity between the new case and stored cases \cite{liao1998similarity}. Furthermore, a machine learning approach must be able to search for similar cases effectively and time efficiently \cite{aamodt1994casebased}. Uncertainty needs to be taken into account, as the available information from previous cases may be incomplete, imprecise, or subjective \cite{weber2007fuzzy, low2019multiple}. It is also beneficial to develop transfer learning techniques that allow recognition of useful knowledge from other systems, and support their integration into the CBR model \cite{aha2009casebased}.

\subsection{Assistive technologies, quantified self, and wearable health devices}\label{subsec:assistivetechquantifiedself}
The robust pace of advancements in energy-efficient computing and sensor miniaturisation \cite{tricoli2017wearable} has enabled wearable devices to become increasingly more capable as assistive technologies. Continuous use of wearable health devices throughout the day can facilitate novel data-driven paradigms in healthcare and personal health management by providing real-time, comprehensive data from sensors to enable personalised monitoring, e.g., fall detection (see Box \ref{box:assistivetechnologies}), early detection of health issues, and tailored interventions \cite{lu2020wearable}. In particular, this technology has led to a rapidly expanding ``quantified self'' movement, where individuals use self-tracking to improve their health and well-being \cite{feng2021how}. The corresponding data, although potentially longitudinally extensive, is similar to rare disease situations, as it pertains to a single individual. In particular, the devices often do not transfer all recorded data to centralised servers for combined analysis with data from other individuals. Instead, the recorded data primarily stays on-device and correspondingly analysis is performed there, e.g., in an edge computing solution \cite{covi2021adaptive}. If the device at least intermittently has access to external servers, these can be leveraged for additional information, e.g., additional data and/or trained models hosted by the manufacturer. However, it is still desirable to analyse the small data on-device, and use this external information as additional context to improve predictions made from the small dataset \cite{wang2020convergence}, as these predictions should still ideally be personalised to the user. This means that some matching to subgroups represented by the external data is needed, including assessment of similarity and potential approaches for information transfer. Much like above, this setting is similar to rare disease applications, yet with the additional complication of having the different sources of information in different locations. On the device itself, the model based on the larger amount of information can then be adapted to the user, to reduce the prediction uncertainty \cite{ferrari2023deep}.

\subsection{Data minimisation}\label{subsec:dataminimisation}
Data minimisation, i.e., the collecting and processing of only the data required for a specific purpose, is a crucial concept whenever dealing with data from individuals. It is governed by regulations such as the EU General Data Protection Regulation (GDPR) \cite{europeanparliament2016regulation}, which also underscores the necessity of informed consent and the need for explainable decision-making models \cite{dexe2022explaining}. In settings where data collection, storage, or transfer should be limited for ethical, legal, privacy, or cost reasons, small data methods can enable data minimisation by preserving useful inference from deliberately limited information.

Data minimisation requires explicit assessment of what information is needed for the task, what uncertainty is introduced by collecting or using less data, and whether missing information can be compensated for by transfer from related data sources, prior knowledge, or pre-trained models. 
When an informative relevant subset can be identified, selecting this subset may be preferable to using all available data. However, this subset may itself be small, the relevant variables may be difficult to identify in advance, or additional collection may be undesirable. Small data techniques are useful in these settings because they provide tools for similarity assessment, information transfer, and uncertainty quantification.

Small data methods can therefore support responsible data minimisation by allowing less data to be collected, copied, or centralised while making the resulting uncertainty explicit. This can reduce privacy intrusions and the risk of unintended data leaks \cite{sanchez-rola2019can, kosinski2013private} (see Box \ref{box:trustedpersonalisedrecommendations} for a hypothetical example). Minimisation enabled by small data methods may additionally be beneficial because collection and analysis of smaller data is often cheaper and can be performed with higher quality \cite{faraway2018when}.

\subsection{Generative AI}\label{subsec:generativeAI}
LLMs, such as that behind ChatGPT, are an example of foundation models, the latest generation of generative AI approaches which provide general representations learned from large-scale datasets \cite{bommasani2022opportunities}. Correspondingly, they are adaptable for a wide range of creative tasks, such as generating artwork \cite{ramesh2021zeroshot, ramesh2022hierarchical} or music \cite{huang2020pop, yang2024uniaudio}, as well as practical tasks such as data augmentation, text generation, and image synthesis \cite{wu2023visual, xu2023mplug2, ko2023largescale}. They can also be applied to fields such as biology or medicine \cite{hou2024assessing, peng2024largelanguage, chang2024bidirectional}. Once tailored to a specific scenario, such foundation models can be used to simulate and explore hypothetical situations, even with limited real data available. This ability to generate synthetic data based on a learned structure, combined with expert insights or established knowledge from a specific field, is particularly attractive in small data settings, as it allows for the conversion of additional knowledge into data.

Yet small data problems still arise, as even large-scale datasets have areas that are sparsely covered, e.g., as illustrated in \ref{box:genAI}. When an LLM is used to generate text pertaining to a part of the underlying statistical distribution that is only supported by limited data, it might also have a tendency to hallucinate, which means it fabricates a plausible but incorrect answer \cite{mckenna2023sources}. This setting, where the text used to prompt the LLM corresponds to the small data at hand, while the LLM represents implicit knowledge from a large-scale dataset, is an example of a more general class of small data settings where a small dataset is used as context information for in-context learning with foundation models \cite{brown2020language}. Therefore, the challenge is to assess how close the small data at hand is to the implicit knowledge represented by the foundation model to detect potential issues with information transfer, such as hallucination in LLMs. Alternatively, foundation models can be fine-tuned to the small data at hand by adapting the model parameters. However, this might be challenging with limited data. Even if such an adaptation is not feasible or straightforward, a model should at least be adapted to warn the user when the provided response is based on an underrepresented knowledge area. The importance of acknowledging and communicating underlying uncertainties in models should be considered when formulating policies that rely on predictions or insights from foundation models, ensuring there is clear communication from developers on uncertainties and limitations in the model predictions and that policies account for potential biases, gaps in knowledge and model limitations.

\begin{myfloat}[tb]
\begin{center}
    \label{box:incontextlearning} 
    \begin{tcolorbox}[colback=similaritycolor, colframe=lightgray,  width=1.0\textwidth, sharp corners, boxrule=0mm]
    \textbf{Glossary: \textit{In-context learning}} allows large language models (LLMs), and more generally foundation models, to perform new tasks without retraining by providing task-specific context information, e.g., prompt text that includes examples, as part of the input during inference. Unlike traditional learning methods that require extensive fine-tuning, in-context learning enables LLMs to adapt dynamically to new problems by interpreting the provided context. For example, when tasked with translating a sentence or generating a summary, the model can infer an appropriate approach based on the information contained in the query itself. This capability has significant implications for rapid prototyping and adaptability in real-world applications.
    \end{tcolorbox}
    \vspace{-1em} 
\end{center}
\end{myfloat}

\section{Methods for working with small data}\label{sec:methodsworkingsmalldata}

\subsection{Contributions of different disciplines}\label{subsec:contributediffdisciplines}

Small data problems occur in a broad range of diverse application fields, as illustrated by the examples presented in Box 1-4 and in Section \ref{sec:exemplaryapplications}. This diversity in applications indicates an increasing recognition of the need to develop specific small data approaches that can be used across disciplines. Currently, since different small data application scenarios are typically addressed by different disciplines, such as computer science, mathematics, or statistics, the methodology developed to address small data challenges is highly diverse. As a result, small data research is currently scattered across disciplines, impeding the realisation of its full potential. 

A particular challenge is the limited exchange of methods between different data science disciplines. For example, mathematics and statistics have traditionally been regarded as the disciplines for small data challenges. Research there has historically been performed in the context of applications with strong assumptions, which are used to gain mechanistic insight and to compensate for small datasets \cite{faraway2018when}. For instance, in infectious disease modelling, assumptions regarding disease transmission and recovery rates are employed to predict the spread of diseases like influenza or COVID-19 when real-time data is scarce \cite{metcalf2017opportunities}. Similarly, in tumour growth modelling, assumptions about the growth rates and limitations of resources within the tumour are used to estimate the progression of cancer and optimise treatments with limited data \cite{alarcon2003cellular}. 

Yet, the computer science community has also recently developed important small data techniques. These approaches are typically based on AI, and known under labels such as transfer learning, few-shot learning or meta-learning, but are only slowly finding their way into other data science communities \cite{friedrich2022there}. In particular, many recent innovations in generative AI have considerable potential for small data challenges, but are currently scarcely discussed within this context. Conversely, while the value of incorporating assumptions and domain knowledge into AI approaches is increasingly recognised (e.g., \cite{kather2022medical, dash2022review, nie2025data, darwin2021pooling}), statistical and mathematical modelling expertise is only slowly being incorporated in AI approaches developed in computer science. 

\begin{myfloat}[tb]
\begin{center}
    \label{box:fewshotlearning} 
    \begin{tcolorbox}[colback=similaritycolor, colframe=lightgray, sharp corners, width=1.0\textwidth, boxrule=0mm]
    \textbf{Glossary: \textit{Few-shot learning}} refers to a method in AI where models are trained to perform tasks using only a small number of labeled examples. This technique is especially useful in scenarios where gathering extensive labeled datasets is impractical, such as rare medical conditions or highly specific tasks in industrial applications. By leveraging knowledge from previously seen tasks or datasets, few-shot learning enables models to generalize effectively, bridging the gap between data-rich and data-scarce environments. 
    \end{tcolorbox}
    \vspace{-1em} 
\end{center}
\end{myfloat}

\begin{myfloat}[tb]
\begin{center}
    \label{box:metalearning} 
    \begin{tcolorbox}[colback=similaritycolor, colframe=lightgray, sharp corners, width=1.0\textwidth, boxrule=0mm]
    \textbf{Glossary: \textit{Meta-learning}}, often referred to as ``learning to learn'', trains models to adapt to new tasks using prior knowledge from multiple datasets. Instead of building a single model optimized for a specific task, meta learning equips the model with a generalized capability to handle diverse problems. This approach is particularly relevant in small data contexts, where the ability to leverage knowledge from related domains can significantly enhance performance. 
    \end{tcolorbox}
    \vspace{-1em} 
\end{center}
\end{myfloat}

Small data applications typically also require strong domain knowledge, but domain experts often do not have access to state-of-the-art small data methods. Even though such connections between disciplines are increasingly recognised and needed, they are often obscured by the lack of a common language \cite{faes2022artificial}. This is particularly relevant with respect to small data methods, as no discipline has yet developed a comprehensive small data framework. Clarifying relevant terminology could remove barriers and serve as a first step towards such a shared framework and language across disciplines. 

To achieve this, it is essential to reflect on the contributions of the different disciplines to current small data methods and how their perspectives complement each other. Firstly, we will present relevant small data approaches used in different communities, including corresponding terminology. We begin by giving a brief overview of the contributions of disciplines and methods developed there to the small data themes similarity, transfer and uncertainty outlined in Section \ref{subsec:commonthemes}, before going into more detail on selected specific classes of methods and approaches.

\subsection{Similarity}\label{subsec:similaritymethods}
Quantifying and incorporating similarity into modelling has traditionally been dealt with extensively in the statistics community. For example, approaches such as matching \cite{austin2007performance} and re-weighting individuals \cite{austin2015moving, borgan1995methods} are used to make groups more comparable, or to selectively integrate information from related groups when focusing on a specific subgroup \cite{weyer2015weighting,richter2019modelbased}. 
The concept of localised regression \cite{loader1999local} allows researchers to tailor models to individuals or subgroups, and can be extended to more flexible nonparametric approaches \cite{samworth2012optimal}. Further, approaches for quantifying the similarity between fitted models \cite{liu2009assessing,dette2018equivalence} can be used to compare different groups, e.g., in clinical time-to-event analyses involving small patient populations \cite{binder2022similarity, binder2022data}. Dissimilarity between groups is also explicitly considered in meta-analysis techniques developed in clinical biostatistics, which combine evidence across multiple clinical trials while accounting for between-study differences.
Differences at the dataset level can also be taken into account by re-calibrating existing models to local conditions \cite{moons2012risk}. Methods for taking into account heterogeneity due to dissimilarity have also been developed for prediction tasks \cite{dejong2021developing}, and there have been suggestions for quantifying similarity in such settings \cite{klein2011characterizing, huang2014similarity}. 

From a computer science perspective, similarity can also be learned directly from data, called metric learning \cite{ren2023visual, mathisen2020learning, wu2013learning, liu2015similarity}. Neural networks can also be used to learn complex representations (latent spaces), in which similarity can be quantified (e.g., \cite{gronbech2020scvae, hackenberg2022deep}). For incorporating similarity at the dataset level, so-called dataset meta features, i.e., summary characteristics describing entire datasets, can be leveraged in \textit{meta-learning} approaches \cite{jomaa2021dataset2vec}.

\subsection{Transfer}\label{subsec:transfermethods}
Approaches for information transfer have been primarily driven by the computer science community and will be presented in more detail in the following subsections. They include techniques for transferring information from trained models via pre-training and fine-tuning strategies. In the context of LLMs, approaches incorporate techniques for \textit{in-context learning} and retrieval-augmented generation. More generally, many information transfer approaches rely on so-called \textit{representation learning}, where models first learn a reusable representation from data that can then be applied to new tasks. This includes techniques for \textit{few-shot learning}, where a model is trained to perform tasks with very limited training examples. 

\begin{myfloat}[tb]
\begin{center}
    \label{box:representationlearning} 
    \begin{tcolorbox}[colback=similaritycolor, colframe=lightgray, sharp corners, width=1.0\textwidth, boxrule=0mm]
    \textbf{Glossary: \textit{Representation learning}} focuses on transforming raw input data into compact, meaningful representations that capture the essential features and patterns within the data. These representations make it easier for downstream tasks, such as classification or clustering, to identify underlying relationships. Techniques like autoencoder deep neural networks are commonly used to uncover these representations, which are particularly valuable in small data settings. For example, in image recognition, representation learning can distil critical features like edges and textures into an efficient format, enabling accurate predictions with limited training samples.
    \end{tcolorbox}
    \vspace{-1em} 
\end{center}
\end{myfloat}

Another strategy to transfer information is to incorporate explicit knowledge, such as physical laws, biological mechanisms, or domain rules. Such approaches are developed in the field of \textit{neuro-symbolic AI}, which aims to integrate explicit knowledge, such as mathematical equations or logical rules into AI models. More generally, approaches for combining knowledge-driven and data-driven modelling allow AI models to leverage such explicit knowledge, e.g., encoded in differential equations.

\begin{myfloat}[tb]
\begin{center}
    \label{box:neurosymbolicai} 
    \begin{tcolorbox}[colback=similaritycolor, colframe=lightgray, sharp corners,  width=1.0\textwidth, boxrule=0mm]
    \textbf{Glossary: \textit{Neuro-symbolic AI}} combines the pattern recognition capabilities of neural networks with the structured reasoning of symbolic AI. This hybrid approach aims to address the limitations of purely neural or symbolic systems by integrating data-driven learning with explicit knowledge representation and reasoning. Neuro-symbolic AI enables systems to reason logically about their predictions while handling complex, unstructured data like images or text.
    \end{tcolorbox}
    \vspace{-1em}
\end{center}
\end{myfloat}

\subsection{Uncertainty}\label{subsec:uncertaintymethods}
Uncertainty lies at the core of mathematical and applied statistics, e.g., when using confidence intervals to quantify uncertainty of model estimates.
Quantifying uncertainty is also a crucial aspect of (network) meta-analysis \cite{nikolakopoulou2016planning}. The need to quantify uncertainty also extends to model parameters in knowledge-driven approaches, such as differential equations used for incorporating domain knowledge \cite{kreutz2018easy}. 
Beyond parameter estimation, uncertainty also arises in model prediction and in the model optimisation process itself, including training strategy and hyperparameter tuning \cite{franke2020neural, wenzel2020hyperparameter, bischl2023hyperparameter}. The latter is of particular relevance in AI-based models, which heavily rely on such hyperparameters to optimise performance. An important class of approaches for reducing uncertainty relies on learning across many datasets, for example through ensemble methods and meta-learning.

\subsection{Foundation models}\label{subsec:foundationmodels}
As introduced briefly in Section \ref{sec:smalldatapolicy}, \textit{foundation models} and in particular LLMs have opened a new era of generative AI, as they provide powerful representations adaptable to a wide range of tasks and applications \cite{bommasani2022opportunities}. Typically, they have many parameters (billions or even hundreds of billions) and are first trained on large datasets to learn a general representation. This is referred to as pre-training, while the process of subsequently adjusting the model to a task-specific, potentially smaller target dataset of interest is called fine-tuning. Such a pre-training and fine-tuning strategy is a prototypical example of information transfer and can be particularly useful in small data settings, where only limited task-specific data are available. 

To avoid the need for large amounts of manually annotated training data, foundation models are often trained in a self-supervised or weakly supervised manner \cite{yuan2023power}. In self-supervised training, labelled samples are derived from the input data itself, e.g., by predicting missing parts or reconstructing perturbed inputs \cite{hendrycks2019using, wang2023scientific}, while weak supervision uses noisy or partially labelled samples that can be generated in an automated way \cite{wang2023scientific}.

Foundation models are typically based on a transformer neural network architecture \cite{vaswani2017attention}. Transformers have originally been developed for natural language processing (NLP) tasks, such as sentiment classification, text summarization, question answering or translation. They leverage the concept of self-attention to learn the patterns that underlie given input sentences. Transformers can be adapted to generative tasks through mechanisms such as autoregressive decoding \cite{stern2018blockwise}, where the model generates one word at a time conditional on previously generated words. 
This principle underlies GPT (generative pre-trained transformer) models, which form the backbone of systems such as ChatGPT \cite{radford2019language, wu2023brief}. Another prominent transformer-based language model is BERT (bidirectional encoder representations from transformers) \cite{devlin2019bert}, which has also been adapted beyond language tasks, such as for protein structure prediction \cite{chowdhury2022singlesequence}.

Foundation models are not limited to text, but have found their way into a wide range of data types and applications, including image \cite{han2022survey, khan2022transformers} and audio data \cite{huang2020pop, yang2024uniaudio}, and multimodal settings \cite{wu2023visual, xu2023mplug2, ko2023largescale}. Well-known examples are image generators such as DALL-E \cite{ramesh2021zeroshot, ramesh2022hierarchical}, where images are generated based on a text prompt entered by the user. Notably, transformer-based foundation models even enable the generation of tabular data, such as data collected from individuals, e.g., customers, study participants, or patients. For example, prior-data fitted networks \cite{muller2022transformers} leverage a transformer architecture for training on a large number of synthetic tabular datasets and subsequently perform inference on a potentially small real-world target dataset without re-training \cite{muller2022transformers, hollmann2022tabpfn, hollmann2025accurate}. Such foundation models for synthetic data are particularly relevant in small data settings, as they use synthetic data as prior knowledge and thereby reduce the amount of real training data required, in line with data minimisation principles.

\subsection{Representation learning, in-context learning, and few-shot learning}\label{subsec:repincontextfewshotlearning}
Many generative AI approaches can be understood as forms of representation learning, in which input data are transformed into compact representations that capture key patterns and enable generalisation and data generation. Representation learning is closely linked to transfer learning, as learned representations allow for sharing information across different datasets, tasks, or data sources. For example, such a representation enables information transfer from the dataset used for pre-training to a target dataset of interest, or even between different data modalities, e.g., in multi-modal foundation models.

A powerful learned representation is also central to in-context learning, where a model performs a new task based on a small set of examples provided at inference time as part of the prompt (i.e., the context) \cite{brown2020language}. This capability is particularly prominent in LLMs, where demonstrations of the task are provided to the model as part of the prompt in natural language. This enables LLMs to adapt to novel tasks without fine-tuning, making in-context learning especially attractive for small data settings.

Another strategy for tailoring LLMs to specific tasks is retrieval-augmented generation. Here, a LLM is linked to an external data source or knowledge base to include additional information \cite{lewis2020retrievalaugmented, gao2024retrievalaugmented, chen2024benchmarking}. This enables information transfer between the external knowledge base and the LLM representation, which can help leverage LLMs for small data settings. This concept is tightly linked to principles of neuro-symbolic AI and more generally to combining data-driven and knowledge-driven modelling, as explained in more detail below.

Few-shot learning techniques can be employed when a small target dataset has a similar format to the data used to train a large-scale model. Few-shot learning mimics humans’ ability to learn tasks after having seen only a few samples. The aim is to leverage the representation obtained from a larger dataset to support predictions for a smaller target dataset \cite{ren2023visual}.

Few-shot learning is an attractive tool for tackling small data challenges, as it aims to optimise performance when training data is scarce. For example, research in few-shot learning focuses on how to best extract features from training data that generalise well to subsequent few-shot tasks, i.e., to find an optimal representation for generalisation. Importantly, such representations should allow adaptation to new tasks without forgetting critical structure from previous trainings \cite{wang2021fewshot}. In small data situations, few-shot methods can also be combined with data augmentation or generative approaches, or leverage knowledge from prior learning across domains \cite{sun2021research} to increase the set of training examples. For example, large-scale image databases such as ImageNet \cite{deng2009imagenet} have been used to enable few-shot learning in medical image diagnosis challenges \cite{wang2017chestxray8,shakeri2022fhist,codella2018skin, samiei2019tcga}.

\subsection{Meta-learning}\label{subsec:metalearning}
Meta-learning refers to a class of approaches that enable information sharing across multiple, potentially small, datasets. Such information transfer can occur at different levels of a model, for example via shared hyperparameters, neural network architectures, or initial parameter values. Meta-learning approaches train models jointly on multiple datasets or tasks in order to extract prior knowledge, which enables rapid adaptation to new target tasks with only few examples \cite{finn2017modelagnostic, loshchilov2018decoupled}.
This strategy has been particularly successful for jointly learning prediction models across related tasks \cite{vanschoren2019metalearning, finn2017modelagnostic}. As several datasets are incorporated at once, meta-learning typically requires less fine-tuning on the target dataset \cite{jadon2023overview}, making it particularly attractive for small data. 

However, a necessary condition for meta-learning is that source and target datasets are not too dissimilar. To address this, meta-learning can be combined with similarity techniques for quantifying and incorporating the similarity of datasets, e.g., via dataset meta-features, i.e., characteristics of the target dataset. For example, auxiliary neural networks have been used for generating such dataset meta-features for tabular datasets \cite{jomaa2021dataset2vec}.

\subsection{Knowledge-driven modelling}\label{subsec:knowledgedrivenmodelling}

In many real-world applications, a substantial body of knowledge about the subject at hand may already be available, e.g., the specialist knowledge of domain experts or knowledge encoded in rules, systems, or mathematical equations. Incorporating such existing prior knowledge into modelling can reduce the amount of additional data needed, making knowledge-driven approaches particularly attractive for small data scenarios. 

This idea has been at the core of much of the research in statistics and mathematical statistics, making these disciplines a key contributor to small data methodologies \cite{faraway2018when}. There has been a recent surge in research in these disciplines for infusing such small data expertise into settings where big data techniques need to be adapted to small data settings, complemented by developments in computer science. There, ideas of knowledge-driven modelling are, e.g., reflected in neuro-symbolic AI approaches. These combine data-driven modelling by neural networks with symbolic AI techniques, which rely on explicit knowledge representations and logical reasoning, e.g., using expert systems, rule-based approaches mimicking human expert decision-making, or logic-based models \cite{hitzler2022neurosymbolic, hitzler2022neurosymbolicapproaches, garcez2023neurosymbolic}. 

In contrast to neural networks, which are powerful for learning complex patterns from data but are inherently opaque and difficult to interpret, such symbolic AI techniques can provide greater transparency, explainability, and verifiability and therefore increase trustworthiness. Neuro-symbolic AI aims to integrate these two paradigms and combine their respective strengths to provide hybrid tools that integrate reasoning and learning from data in an efficient and transparent way \cite{sheth2023neurosymbolic, garcez2023neurosymbolic}. Thus, they can provide more explainable AI approaches while enabling the incorporation of prior knowledge into learning \cite{sheth2023neurosymbolic}. These properties may be particularly desirable in sensitive applications where explainability is crucial, such as in precision medicine, or for policymaking in general.

Knowledge-driven modelling is a dominant paradigm in statistics and many areas of mathematics and physics. Here, research often builds on a strong body of prior knowledge, which is typically reflected in strong modelling assumptions, e.g., in statistics, when pre-specifying the structure of a regression model for assessing treatment effects in a clinical trial. Similarly, in physics or systems biology, temporal dynamics are typically modelled by differential equations that are defined based on domain knowledge, e.g., about the relevant interactions and regulatory processes to describe a small set of molecular or cellular quantities \cite{bruggeman2007nature}. To increase flexibility, such knowledge-driven modelling can be combined with more general-purpose data-driven techniques \cite{li2022machine}, resulting in tools that are both more data-efficient and more interpretable. For example, neural differential equations \cite{chen2019neural, kober2022individualizing} integrate mechanistic knowledge into neural networks by combining them with differential equations, thus directly combining data-driven and knowledge-driven modelling. More generally, neural networks can replace or augment components of mechanistic models that cannot be fully specified \cite{rackauckas2021universal, lagergren2020biologicallyinformed}, an approach often implemented using differentiable programming frameworks that integrate modelling components from machine learning and scientific computing \cite{innes2019differentiable}.

When domain knowledge can be expressed as prior distributions over model parameters, statistical modelling provides a rich set of Bayesian techniques for model building and evaluation \cite{vandeschoot2021bayesian}. Yet, when such a precise specification is not feasible, but knowledge still is available in the form of models, there are other ways for incorporating such knowledge, e.g., through data augmentation, domain adaptation, or regularisation \cite{ba2019blending, gongora2019applications, raissi2018deep}. Pre-existing knowledge can also be incorporated based on expert input in the form of probabilistic relationships of variables \cite{vonrueden2022informed}. These approaches are often grouped under the label ``informed machine learning'' \cite{vonrueden2023informed}, with domain-specific variants such as physics-informed learning \cite{karniadakis2021physicsinformed}.

\subsection{General challenges}\label{subsec:generalchallenges}

\subsubsection{Overfitting}\label{subsec:overfitting}
Training models on limited data incurs the risk of overfitting, i.e., when the model has learned characteristics unique to the limited training data and is aligned too closely to it, but does not generalise to other datasets. While overfitting can also occur in large datasets, small data settings are particularly prone to it due to the limited information available. With fewer data points, there may not be enough diversity to cover a wide range of situations, making it more likely that the model will pick up noise or specific patterns that do not generalise \cite[p. 108--114]{goodfellow2016deep}. 

A related problem can occur when the observations fall into several categories, and some categories are represented more often than others. For example, in a longitudinal clinical dataset on a severe chronic disease, there might be more patients with comorbidities than without, i.e., the presence of comorbidities is overrepresented. The model is then biased towards predicting the more prevalent class, leading to overrepresentation, or may fail to predict the minority class, leading to underrepresentation \cite{rahman2013addressing}. In small data settings, there may simply not be enough data points within a singular class in order to properly assess it.

When models are based on historical data, that is, data that was collected in the past, there is no guarantee that this data will remain in this form over time, or is the same in other datasets \cite{vollmer2020machine}. Potentially necessary adaptation could again be based on approaches for similarity and transfer.

\subsubsection{Validation}\label{subsec:validation}
When developing a model on a specific dataset of interest, the data is typically split into a dataset for training the model and a second dataset for testing the model \cite{ucar2020effect}. Sometimes, when training the model, data from the test dataset is accidentally used during the training process. This phenomenon is known as data leakage. While this is an issue for both big and small data, it tends to have more pronounced and unpredictable effects on small datasets \cite{rosenblatt2024data}. Because the datasets are so small, this can lead to an overestimation of model accuracy, a decrease in its ability to generalise to new data (known as generalisability), and the introduction of bias.

After internal validation, models need to go through external validation to see whether the results are generalisable to other, similar datasets \cite{li2024machine}. In a small data scenario, this might be rather challenging to realise if there is no sufficient external data available, or if only dissimilar data is available. Assessment and incorporation of similarity is therefore crucial. To remove hurdles, policies should facilitate streamlined data exchange and recommend the creation of data in accordance with the FAIR (Findable, Accessible, Interoperable, Reusable) principles \cite{wilkinson2016fair} to enable joint use of similar datasets for validation, or already for model development.

\section{Conclusions and future perspectives}\label{sec:conclusionsperspectives}
In an increasingly data-centric world, it is important to make full use of new technologies, such as AI techniques. As shown, the potential benefits are not limited to big data settings, but in particular small data settings can benefit and also need more attention, e.g., to adequately serve underrepresented groups. Novel data analysis and modelling techniques for small data settings can then complement technical developments, such as in assistive technologies, to realise their full potential.

We have provided a variety of case studies and application settings not only to highlight the broad applicability of small data solutions, but also to show the potential from areas where small data techniques are more established, e.g., for rare diseases, to those where the potential is apparent but techniques still need to be established. 

However, a wide range of applications, each with contributions from many stakeholders and academic disciplines, runs the risk of confusing newcomers and reinventing the wheel. To enable sufficient communication across disciplines, a shared language is needed as a first step towards a joint small data framework. This motivated us to jointly present ideas from different disciplines, such as historical and policy perspectives from social sciences, transfer learning from computer science, and the quantification of similarity and uncertainty from statistics and mathematics. 

As the common themes of similarity, transfer, and uncertainty may be more generally useful for structuring the field of small data methods, we have provided an overview of how these themes are approached by the different disciplines, including links between them. Hopefully, this will stimulate the fusion of knowledge-driven approaches, as traditionally used in statistics and mathematics, and data-driven approaches, as now prominently established with deep learning techniques in computer science. Such a fusion could be the key to unlocking AI techniques for small data settings. As method development in different disciplines is currently rather siloed, further investment may be warranted to gain commitment from different disciplines and allow for the development of common methods, as well as analysis of barriers to interdisciplinary small data research.

As an example of a specific technique that could be considered a starting point for such endeavours, foundation models, such as LLMs, appear to be a particularly promising approach for bringing additional information to small datasets. As described, there is currently a vibrant community developing foundation models for different fields, and training/fine-tuning approaches that can incorporate many different types of data and models. This flexibility could also offer an avenue for contributions from different disciplines, where disciplines focused on knowledge-driven approaches could, for example, provide knowledge in the form of synthetic data that can then be ingested by data-driven approaches. Of course, this will also depend on establishing a shared language and a forum where different disciplines can develop a common understanding in order to make the best use of different expertise.

Developments in the methods community must be accompanied by a strong push to expand its application across more fields. So far, many assume that fields which cannot offer big data are cut off from novel methods, such as AI. Therefore, it is of prime importance to raise awareness of the opportunities of small data. One way this could be achieved is through dedicated initiatives that bring together all stakeholders. This would also be critical for ensuring feedback from the specific application fields into the broader methodological community, for steering the small data field in the right direction, and for realising the full potential of small data in everyday life.

\vspace{0.5cm}

\noindent\textbf{Acknowledgments}

\noindent This work was funded by the Deutsche Forschungsgemeinschaft (DFG, German Research Foundation) -- Project-ID 499552394 -- SFB 1597.
We gratefully acknowledge support from the \href{https://royalsociety.org/}{Royal Society}, which contributed to the successful completion of this research as part of the development of the \href{https://royalsociety.org/news-resources/projects/disability-data-assistive-technology/}{Disability Technology report}.

\printbibliography

\end{document}